# User Driven Functionality Deletion for Mobile Apps

Maleknaz Nayebi
EXINES Lab, York University
Toronto, Canada
mnayebi@yorku.ca

Konstantin Kuznetsov
CISPA
Saarbrucken, Germany
kuznetsov@st.cs.uni-saarland.de

Andreas Zeller
CISPA
Saarbrucken, Germany
zeller@st.cs.uni-saarland.de

Guenther Ruhe
SEDS Lab, University of Calgary
Calgary, Canada
ruhe@ucalgary.ca

*Abstract*—Evolving software with an increasing number of features is harder to understand and thus harder to use. Software release planning has been concerned with planning these additions. Moreover, software of increasing size takes more effort to be maintained. In the domain of mobile apps, too much functionality can easily impact usability, maintainability, and resource consumption. Hence, it is important to understand the extent to which the law of continuous growth applies to mobile apps. Previous work showed that the deletion of functionality is common and sometimes driven by user reviews. However, it is not known if these deletions are visible or important to the app users. In this study, we performed a survey study with 297 mobile app users to understand the significance of functionality deletion for them. Our results showed that for the majority of users, the deletion of features corresponds with negative sentiments and change in usage and even churn. Motivated by these preliminary results, we propose RADIATION to input user reviews and recommend if any functionality should be deleted from an app's User Interface (UI). We evaluate RADIATION using historical data and surveying developers' opinions. From the analysis of 190,062 reviews from 115 randomly selected apps, we show that RADIATION can recommend functionality deletion with an average F-Score of 74% and if sufficiently many negative user reviews suggest so.

## I. INTRODUCTION

It is often assumed that the evolution of a product implies constant addition to it which results in a larger and more complex codebase. This addition has been discussed in terms of evolving software code, enhanced quality, added features and functionalities, etc. over different releases of a product. The tendency to add more and more features to an evolving software is a form of excessive software development [39] and does not automatically make the software better. In particular, release planning as an iterative and evolutionary process has been always concerned with further adding features into the next releases [11]. Lehman's [17] sixth law of software evolution emphasizes growth and states that "the functional content of a program must be continually increased to maintain user satisfaction over its lifetime." However, viewing through the lens of user-computer interaction, when a program is mainly invoked by users, the increasing set of features is in sharp conflict with usability [42]. Mobile apps in particular can seriously suffer from this type of problem [43].

On mobile devices, any functionality comes at a cost: First, the small screen severely limits the number of features that can be offered by an application in each UI [10]. Second, computational demands and memory usage may impact battery life. Hence, developers should have an interest in *removing* functionality that negatively impacts the user experience. While this removal can be the result of different development activities (for example, removing the code, commenting out the code, or disabling respective UI elements), from the user's perspective, a functionality is removed when it is no longer accessible through the user interface [25]. There is an established body of knowledge on the release engineering of mobile apps. Several techniques [20] have been proposed for the release planning of mobile apps. Generally, these existing methods are focused on feedback development planning, based on user reviews. They first categorize reviews into general categories of uninformative comments, feature requests, bug reports, or praise. Then, they aim to satisfy that user feedback in the upcoming release. Palomba et al. [29], [30] proved empirically that mobile app developers are changing their code based on the crowdsourced app reviews. Among these studies, multiple provided a variety of taxonomies for mobile app reviews [6], [32]. When analyzing user reviews, a few studies reported a reason for negative reviews [15], [19]. The study of Nayebi et al. [25] showed that 11.23% of commits and 44.79% of the developers indicated better user experience as the reason for deletion. The author's analysis of commit messages showed that 14.63% of deletions are driven by negative user feedback. Yet, users' perceptions of feature removal have never been evaluated empirically or ever surveyed with the users.

Our research focuses on studying feature deletions in the evolution and release planning of mobile applications and their visibility to the end users. To understand the significance of this issue to users, we conducted a survey of 297 individuals. Since functionality is typically accessed through graphical user interface (GUI) elements [2], we specifically investigated deletions that are visible to end users. By surveying 297 individuals, we examined the extent to which users notice functionality deletions over different releases, their perception and emotional response to such changes, and any resulting alterations to their usage patterns. Driven by the results of this survey, we introduce RADIATION[1], a system that analyses user reviews and recommends UI elements and features that can be considered for deletion. We evaluated RADIATION internally (via cross-validation) and externally (with 37 developers and 42 users). Results show RADIATION recommends feature deletions with high precision (0.83 in retrospect and 0.95 when compared to developers). End-user study confirmed recommendations' validity.

---
[1] RADIATION = **R**eview b**A**sed **D**elet**I**on recommend**ATION**

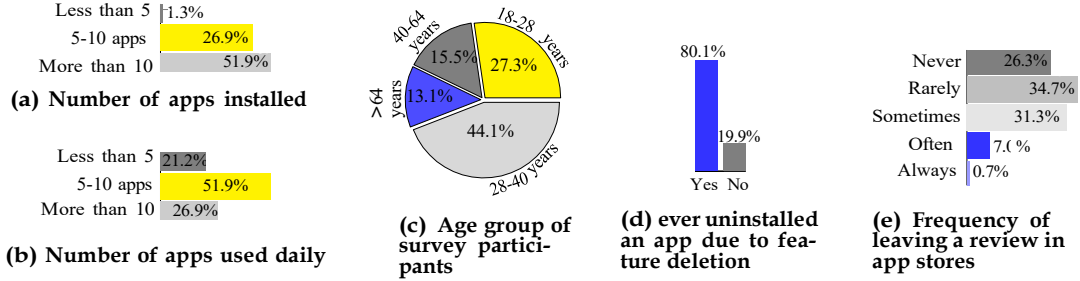

Fig. 1. Demographics of 297 participants in the survey. Regional census categories are used for age information (the region is masked due to double-blind).

## II. IMPORTANCE OF FEATURE DELETIONS TO USERS

To the best of our knowledge, the significance of feature deletion for end users is unexplored. Nayebi et al. [26] examined the problem from the developer's perspective and developed a taxonomy of deletion commits in mobile apps using source code and commit messages. However, this taxonomy covers a broad range of artifacts and reasons, and it is unclear whether and to what extent these deletions are visible or important to end users;

*"How is the deletion of software functionality perceived by mobile app end users?"*

To answer this question and understand the relevance of app feature deletion, we surveyed real app users. We followed the established guidelines for performing the survey research [34]. Our survey consists of four main parts:

- Gather the demographics,
- Assess how aware mobile app users are of missing features or functionalities,
- Evaluate if the deletion of features impacts users' satisfaction, and
- Understand the extent and impact of functionality deletion or limitation on app usage.

The survey included 12 questions overall, and they were all close-ended questions (see Table I). Five questions were designed to capture demographics. The rest of the questions sought participants' opinions using a five-point Likert scale. The survey was focused on individuals' experiences and decisions. The survey was anonymous, and we did not gather any identifying information from the participants. We used `Qualtrics` as the survey instrument.

For acquiring participants, we used convenience sampling [16]. We posted the survey through our personal connections on social media. The link to the survey has been clicked 638 times. 388 individuals started the survey, whereas 297 individuals completed the survey and responded to all the survey questions (46.5% of all the people we could reach). Among the 297 participants, 44.1% were aged between 28-40 years old. 27.3% were 18-28 years, 15.5% were 40-64 years, and 13.1% were above 64 years old. The majority of the participants (51.9%) have personally installed 5-10 apps on their devices[2]. 26.9% (80 participants) have installed more than 10 apps, while 21.2% of the participants have installed less than five apps personally. Also, 53.9% of all the participants (160 individuals) used more than 10 apps on a daily basis. Only 1.3% of participants (only four individuals) used less than five apps daily, while 44.8% used 5-10 apps daily. Out of the 297 participants, 238 individuals (80.1%) have uninstalled some apps but only 39% sometimes or more frequently have left any reviews for a mobile app. The demographics are presented in Figure 1. Our questions followed three main objectives:

**First,** the extent to which a user realizes and notices the change and deletion in mobile app features (Q6 and Q7 in Table I), The majority of the users (55.2%) *sometimes* noticed changes in the app features that they were using. While 2.7% of them (8 out of 297 participants) and 20.5% reported they have *never* or *rarely* noticed a change. When it comes to the deletion of features, 34.4% *never* or *rarely* noticed a deletion.

[2]Mobile devices come with a number of pre-installed apps.

TABLE I
QUESTIONS USED IN SURVEYING MOBILE APP USERS. THEY ARE SHORTENED FOR PRESENTATION PURPOSES.

| ID | Question | Response type |
|---|---|---|
| **User Demographics** | | |
| Q1 | How many apps have you personally installed on your phone, currently? | <5, 5–10, >10 |
| Q2 | How many apps do you actively use daily? | <5, 5–10, >10 |
| Q3 | What is your age group? | Four life groups |
| Q4 | How often do you leave an app review? | Five-point scale |
| Q5 | Have you ever uninstalled an app? | Yes/No |
| **User Realization** | | |
| Q6 | How often have you noticed that functionality you have been using is changed (is different from before) in an app? | Five-point scale |
| Q7 | How often have you noticed a functionality you have used is no longer available? | Five-point scale |
| **User Perception** | | |
| Q8 | How did you feel about the lack of access to the app functionality? | Five-point scale |
| Q9 | To what extent has this now missing functionality impacted your app usage? | Five-point scale |
| **User Decision & Action** | | |
| Q10 | How often did you leave a review for an app following the deletion of a feature? | Five-point scale |
| Q11 | How often did you look for an alternative app to install following the deletion of a feature? | Five-point scale |
| Q12 | How often have you uninstalled an app following the deletion of a feature? | Five-point scale |

This compares to the 65.7% who reported sometimes or more frequently noticing a feature deletion in an app.

**Second,** the perception and sentiment of users toward a feature deletion in an app and its impact on their app usage (Q8 and Q9 in Table I), 51.9% of participants perceived somewhat of *negative* feeling associated with feature deletions. 41.1% of the participants stated negative and 7.75% stated *very negative* sentiments. This is while 13.5% was *positive*, and 1.0% stated *very positive* feelings about feature deletions. 33.7% of the participants were *neutral* about the feature deletion. Almost the same proportion of users (48.8%) reported almost *no change* in their app usage following a feature deletion. Yet, 51.2% reported *somewhat* or *extensive* change in their app usage following a feature deletion.

**Third,** the extent that deletions impact users' decisions and provoke a reaction (Q10 - Q12 in Table I), Only 51 out of 297 participants (17.17%) have *often* or *sometimes* left a review for a mobile app following the deletion of a feature (Q10). This compares to the 39.1% of the participants who *sometimes* or *often* left a review for an app (see Figure 1-(e)). As a result of losing access to app functionality, 63.7% of the participants *sometimes* or *more frequently* looked to use alternative apps. 36.4% of the participants *never* or *rarely* looked up alternatives when their access to a certain feature is omitted (Q11). 31% of the participants reported that they *at least once* uninstalled an app because of a feature deletion. 41.4% *never* or *rarely* deleted an app due to this reason while 27.6% *sometimes* did so (Q12).

Figure 2 shows the summary of our survey results.

> *Deletion of app functionality provokes negative feeling for the majority of the participants (51.9% of the participants) and somewhat change their usage behavior (51.2% of the participants). Functionality deletion caused 31.0% of the users often to migrate to another app. 27.6% of the users uninstalled the app following the deletion of a feature.*

Hence, we consider the issue of functionality deletion visible to the end user. Further, Nayebi et al. [26] stated that while the planning for feature deletion is less frequent compared to the feature additions and bug fixes, still, 77.3% of developers **plan** for these deletions. The desire to retain users and avoid the distribution of negative sentiments about an app motivates us to evaluate if and to what extent these deletions are predictable.

### III. RESEARCH QUESTIONS AND EMPIRICAL DESIGN

Functionality deletions are important to users, and eliminating access to particular features can cause customer churn, negative reviews, and lead to app uninstallations. Hence, these deletions should be planned with care and precision by a software product team. To assist the production team with such decisions, we introduce RADIATION to recommend deletions based on user reviews. We further evaluate RADIATION's performance retrospectively and by performing cross-validation. To externally validate RADIATION, we survey 37 software

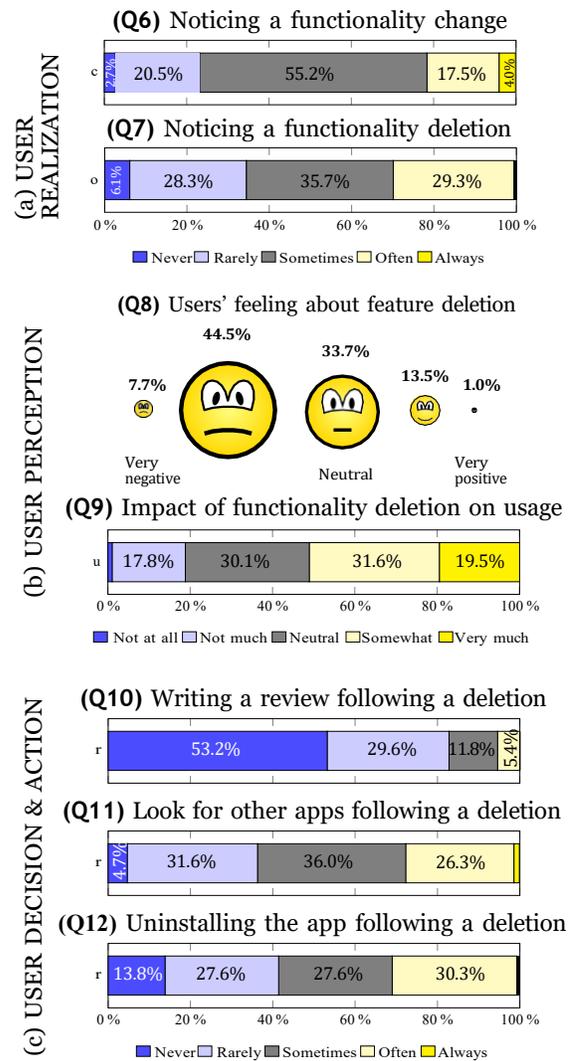

Fig. 2. Results of the survey with app users (Q6 to Q12).

developers and 42 users to understand their perception of the value of deletions recommended by RADIATION. We evaluated RADIATION in three ways by answering the following research questions:

**RQ1:** To what extent does RADIATION accurately predict functionality deletions in comparison to actual deletions, retrospectively?
*RADIATION predicts the elimination of the functionality which is visible to the end user. It connects reviews to the UI elements that represent the functionality as seen by the end user. For the internal validation, we randomly sampled 115 apps and cross-validated the results of RADIATION with the actual changes that happened retrospectively. We gathered deletion commits and manually checked the code base following former studies [26] and compared actual deletions with the RADIATION suggestions.*

**RQ2:** To what extent do app developers consider analogical reasoning useful for predicting functionality deletions?
*We performed a survey with 37 developers to evaluate if the predictions of RADIATION would make sense to the*

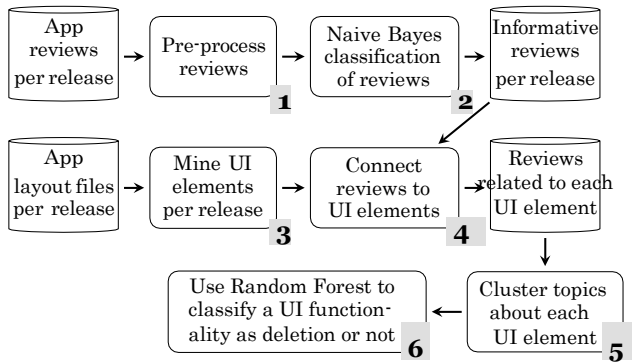

Fig. 3. The process of RADIATION to support decisions on user-driven UI functionality deletions.

*professionals. These developers are active in open-source mobile app development but are not the actual developers of the app. We provide each developer with the reviews, the link to the code repository, the app, and the recommendations of RADIATION and ask if they consider these suggestions reasonable or not.*

**RQ3:** What was users' experience with the functionalities that *Radiation* offers for deletion?
*We conducted a survey with 42 users to assess their sentiment towards the functionalities recommended for deletion by RADIATION. After familiarizing themselves with the app, we asked each participant to evaluate 30 UI functionalities based on their level of liking and the importance of deletion. We performed a controlled experiment by presenting the question for both the features recommended for deletion by RADIATION and those not recommended for deletion. Finally, we analyzed the relationship between user sentiment and the recommendations provided by the tool.*

RADIATION can recommend feature deletions sufficiently well. Our evaluation of RADIATION on 115 apps across 3,364 releases and for 190,062 reviews shows a recall of 0.48 and precision of 0.83 using 10-fold cross-validation (**RQ1**). Our evaluation of 25 apps involving 36,039 reviews with 37 developers shows an F-score of 0.90 for RADIATION (**RQ2**). Also, our survey shows users' negative experience with the features that RADIATION recommends for deletion (**RQ3**).

## IV. RADIATION FOR PREDICTING FUNCTIONALITY DELETIONS

Multiple factors may trigger functionality deletion. We designed RADIATION to recommend deleting functionalities suggested by user reviews. However, since apps may receive a large number of reviews, manually tracking user feedback may not be feasible. The current literature on apps' user needs and planning is primarily focused on adding features or fixing bugs in each release, based on user requests [12], [32], [44]. RADIATION differs from this approach by targeting deletions inputting user reviews. RADIATION is a recommendation tool that helps developers identify deletion candidates. While deleting features is sometimes necessary [26], developers must be cautious about the features they remove, as it can result in a negative user experience and potentially losing customers, as shown by our survey study (see Section II). RADIATION is the first step to assist developers with this task. Figure 3 illustrates the six steps of RADIATION. In what follows, we explain each step of our proposed method and provide a walk through examples referring to Figure 4.

**Step 1. Reviews pre-processing.** We eliminated emojis, special characters, and stop words and expanded contractions ("can't" was expanded to "can not"). Then, we applied lemmatization to map the words into their dictionary format ("deciding" and "decided" turned into "decide"). We used `Python` library `NLTK` for this step. We customized the list of stop words as suggested by Maalej and Nabil [18] and Palomba et al. [29].

**Step 2. Separating informative and non-informative reviews.** Not all reviews were useful. We followed the definition of what is informative and non-informative as described by Maalej and Nabil [18]. In short, informative reviews communicate content that can be used in the process of the app evolution, while an advertisement, a short statement of praise (i.e., "The app is nice"), or a statement of an emotion (i.e, "I hate this app!") is not informative for enhancing an app in future releases. To identify informative reviews, we manually classified a fraction of reviews (see Section V) and used them to train a Naive Bayes classifier (following [18]). This setup resulted in the F1 score (the harmonic mean of precision and recall [35]) of 0.82, calculated as the average of ten 10-fold cross-validation runs.

**Step 3. Finding UI elements for each release.** For *each release* we extracted UI elements used in an application. We leveraged the UI elements to connect the reviews with the apps' functionality following the method of Palomba et al. [29]. They showed that users write reviews related to the app components visible to them, which are the elements of the user interface. To mine UI elements, we implemented the lightweight analysis of Android layout files. These files include most of the GUI elements, also known as `view widgets`, and control as it is visible to the app user [1], [21]. Additionally, we parsed the `Strings.xml` file which contains text strings for an app. By mining these files, for each identified UI element we got its *description* consisting of an element type, a variable name used in the code, a label associated with the element, and an icon name if applicable (e.g., <Button, btn_mic, 'Start Listening', ≥).

**Step 4. Connecting reviews to the UI elements.** We used the description of elements connecting reviews to app functionalities. To connect a review to a UI element in a release $V_i$, we calculated the cosine similarity between the text of a UI description and a review's content. We established connection when the similarity score exceeded a threshold of 0.65. Palomba et al. [29] used the threshold of 0.6 for this purpose, however when analyzed manually, we slightly increased the threshold to achieve a more accurate matching.

**Step 5. Clustering reviews based on their topic.** Several app reviews are pointing to the same functionality, while they may contain different opinions about that functionality.

We used *Hierarchical Dirichlet Process* (HDP) [41] with its default setup to group reviews related to each functionality (UI element) as suggested by Palomba et al. [31]. HDP is a topic mining technique which automatically infers number of topics. Using HDP as described in [31], we performed topic modeling and formed clusters with reviews about a particular topic. One review might also discuss multiple UI elements hence the clusters are non exclusive. We manually analysed the results for 1,500 reviews across eight apps: The topics were intuitive and understandable.

**Step 6. Identifying candidate functionality deletion.** Following the existing literature on prioritizing app reviews (Table II) and our survey (Section II) we selected attributes for identifying and recommending possible functionality deletion. To determine candidates, we used Random Forest, as it was suggested by related studies [44] and showed good time performance. A list of attributes for training is presented in Table II. The "polarity" and "objectivity" of the reviews in a cluster were extracted by sentiment analysis performed by Pattern [23], [24], [37], [40] technique. We evaluated the classifier based on 190,062 reviews across 115 randomly chosen apps.

Figure 4 illustrates the execution of RADIATION on the WIKIPEDIA Android app.

## V. EVALUATION AND CASE STUDY DESIGN

As of June 2022, F-Droid (the open-source repository for Android mobile apps) included 3,810 mobile apps. Among them, we identified 1,704 apps with a valid link to their `GitHub` repositories. These apps involve an overall of 14,493 releases. As deletions are identified by comparing sequential releases, we excluded 554 apps which had only one or two releases from our analysis to evaluate RADIATION over multiple releases. For the remaining apps, we gathered the reviews from the `Google Play` store while accessing their code and development artifacts through `GitHub`.

We randomly selected 8,300 reviews ($\cong$ 5% of the total number of reviews) across different apps and manually labeled each review as "informative" or "non-informative" as described in Step 2 of RADIATION. Two of the authors classified these reviews with an average Cohen's Kappa agreement's degree [38] of 86%. We labeled 2,917 of these reviews as "non-informative" and used them along with the same number of "informative" reviews randomly sampled from the rest of reviews to train a classifier. Finally, we identified 8.1% of the total number of reviews as uninformative.

We applied RADIATION and analyzed 115 randomly selected apps in detail as well as evaluating RADIATION recommendation against developers judgment (**RQ2**) and users experience (**RQ3**) for 25 apps. In what follows, we explain the methodology for answering each research question and then provide the results.

### A. Internal Validation of RADIATION (RQ1)

To internally validate the usefulness of RADIATION, we retrospectively compared the recommendations of RADIATION with the actual changes in the source code. We performed this cross-validation across multiple releases of the same app and for a total of 115 apps, involving 3,364 releases.

As a result of Step 5 of the RADIATION process, we clustered the reviews for each UI element. Next, we manually labeled each review cluster as either "deleted" or "not deleted". This labeling was conducted by two independent researchers who manually checked for the deletion of the code in the source code repository and identified the deletion commit messages, as discussed in the literature [26]. The agreement between the annotators was close to perfect, with a 96% agreement rate, as the decision was based on factual evidence of changes in the Git repository. Any differences were resolved with a short code look-up and recheck. Hence, if an element $E_i$ was deleted in release $V_i$, we tagged the clustered reviews in $V_{i-1}$ as "deleted". We used these manually labeled clusters as our *truth set*. To internally validate our results, we compared the output of RADIATION with this truth set. RADIATION takes the information of the app (as detailed in Table II) in release $V_{i-1}$ and predicts whether an element $E_i$ in release $V_i$ should be deleted or not. Retrospectively comparing this prediction with our truth set can result in one of the following cases:

**TP:** RADIATION recommends deletion of $E_i$ in $V_i$, and historical data of our truth set shows the element was deleted.

**TN:** RADIATION does not recommend deletion of $E_i$ in $V_i$, and historical data of our truth set shows the element was not deleted.

**FP:** RADIATION recommends deleting $E_i$ in $V_i$, but our truth set's historical data shows that the element was not deleted.

**FN:** RADIATION does not recommend deletion of $E_i$ in $V_i$, but historical data of our truth set shows its deletion.

Using these outcomes, we formed a confusion matrix and calculated the precision, recall, and F-Score of RADIATION.

TABLE II
FEATURES USED IN RF TO RECOMMEND IF SOME FUNCTIONALITY IS A CANDIDATE FOR DELETION.

| Attribute | Reason | Description |
|---|---|---|
| $\|Reviews\|$ | [5], [44] | The number of reviews in a cluster. |
| $rating$ | [5], [44] | Each app reviews is associated with a rating. $rating$ is the average rating of reviews in a cluster. |
| $\Delta\ rating$ | [44] | $\Delta$ between the average rating of the cluster and the average rating of the app in a specific release. |
| $\overline{polarity}$ | [12], [32], [33] | The average polarity of the reviews in a cluster. Polarity is one dimension of sentiment and is a number between [−1, 1]. −1 shows negative sentiment, 0 is neutrality, and 1 is the very positive feeling. |
| $\overline{objectivity}$ | [26] & our user survey | Average objectivity of the reviews in a cluster. Objectivity is another aspect of sentiment and is a number between [0, 1]. 0 shows the message is totally objective (expression of facts) and 1 shows the message was opinionated (subjective) [40]. |
| $\|uninstall\|$ | [26] & our user survey | The number of reviews talking about "uninstalling the app or requesting "refund". |

**Step** 1 . **Pre-processing of reviews:** The result of this step is a lemmatized and cleaned set of reviews. This way, for example, a review such as "I can't use save pages as it keeps crashing" became "I can not use save page keep crash".

**Step** 2 . **Filtering:** Non-informative reviews are eliminated as the result of machine learning classification. For instance, a review such as "I hate this app!" or "The app is awesome" is non-informative.

**Step** 3 . **Collecting UI Elements in release V2.0:** Share via, menu, *saved pages*, location service, close all tabs, ...

**Step** 4 . **# of reviews associated to *saved pages*** = 71

| Review | Associated element | Release |
|---|---|---|
| Updates deleted my saved pages in the offline mode | Saved pages | V1.8 |
| I can't share anything on Facebook or Google+ | Login | V2.0 |
| In offline smd airplane mode I can't view a saved page | Saved pages | V2.0 |
| Saved pages doesn't sync. between my devices | Saved pages | V2.0 |
| ... | ... | ... |

**Step** 5 . **Clustering reviews related to *saved pages* by topic:** Offline option, Saved pages.

**Step** 6 . **Recommending deletion using Random Forest:** Attribute values related to clustered reviews:

| Reviews topic | polarity | \|Reviews\| | rating | objectivity | \|uninstall\| | Δ rating |
|---|---|---|---|---|---|---|
| Offline option | -0.34 | 24 | 3.28 | 0.15 | 0 | 0.92 |
| Saved pages | -0.41 | 47 | 1.32 | 0.22 | 2 | 2.88 |

Example of a regression tree generated from the Random Forest classifier:

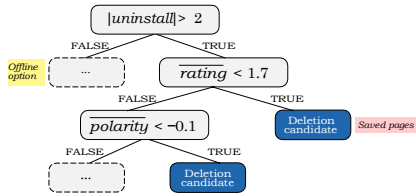

**Recommendation:** *Saved pages* is a candidate for deletion.

**Evaluation** Evaluating HDP topics with developers:

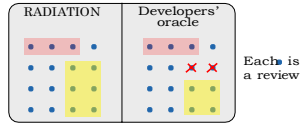

**Retrospective analysis:** We compared V2.0 and V2.1 and found that Δ(V2.0,V2.1) = *Saved pages*. *Saved pages* was deleted.

Fig. 4. An example of RADIATION process on WIKIPEDIA app

For this evaluation (**RQ1**), we excluded apps with less than two releases (554 apps). Among the remaining 1,150 apps, we picked 10% (115 apps) randomly and analyzed them in depth. These 115 apps included 190,062 reviews.

### B. External Validation of RADIATION with Developers (RQ2)

We aimed to evaluate the perception of software developers regarding the correctness of the RADIATION recommendations. Initially, we invited software developers who actively commit to the repositories of our studied open-source `Android` apps to participate in the study. However, due to their unavailability and unresponsiveness, we decided to recruit developers through advertising on our social media and professional network. We specifically targeted developers to participate in a survey on app functionality deletion. Using convenience sampling, we were able to hire 37 developers for the study. These developers had an average of 8.3 years (ranging from two to 15 years) of experience in software development and 4.4 years of mobile app development (ranging from one to 12 years). Each of the developers had participated in the development of at least two apps. In evaluating RADIATION, the developers went through two steps: topic modeling in Step 5 , and reviewing RADIATION recommendations for 25 apps and 36,039 reviews (20% of our chosen apps for validation).

*1) Evaluation of cluster topics about each UI element:* The quality of topics and modeling in Step 5 is crucial to the success of RADIATION. To assess the effectiveness of clustering by HDP in Step 5 of RADIATION, we utilized a human judgment method called *topic intrusion* [4]. This involved presenting the top two topics with the highest similarity for a review and presented them along with a random topic of lower probability (the intruder topic) to a developer, who was then asked to identify all relevant topics. To evaluate the results of Step 5 we calculated *Topic Log Odds (TLO)* [4].

TLO is a quantitative measure of agreement between a model and a human. TLO is defined as the difference between the log probability assigned to the intruder topic and the log probability assigned to the topic chosen by a developer. This number is averaged across developers to get a TLO score for a single document $d$ [3]:

$$TLO(d) = \frac{\sum_s \log \theta_{r,trueintruder} - \log \theta_{r,intruderselectby's'}}{S}$$

Where $\theta_{r,t}$ is the probability that a review $r$ belongs to a topic $t$, and $S$ is the total number of developers.

*1) Evaluating RADIATION recommendations with developers:* Further, to evaluate the RADIATION recommendations externally, we provided the cluster of reviews about a UI element (the output of Step 5 ) and asked developers to categorize each cluster as either "motivating functionality deletion" or "not motivating functionality deletion". We then compared the results of RADIATION with developers' opinions to evaluate its performance. As a result of this comparison, one of the four cases of TP, TN, FP, and FN could occur. However, unlike the cases discussed in Section V-A, here we compared the RADIATION recommendations with the developer's judgment rather than the historical data. In consideration of the number of participants in our survey, we randomly selected 25 apps and had each functionality cluster evaluated independently by three developers and got final decisions by majority.

**LOOP HABIT TRACKER**

Do you need more time to familiarize with the app? ○Yes ●No
Q1- How did you like the "Detailed Scoring of Daily Progress" feature?
○Strongly dislike (-2) ●Dislike (-1) ○Neutral (0) ○Liked (+1) ○Strongly Liked (+2)
Q2- How do you feel if the "Detailed Scoring of Daily Progress" feature is being deleted?
○Strongly disappointed(-2) ○Disappointed (-1) ○Neutral (0) ●Liked (+1) ○Strongly Liked (+2)

Fig. 5. Sample questions asked for evaluating RADIATION with users.

## C. External Validation of RADIATION with Users (RQ3)

We aim to assess the degree to which recommendations generated by RADIATION align or conflict with user experience toward specific app functionalities. As part of our evaluation, we randomly selected 30 UI elements and functionalities from each app. We made a deliberate effort to include a mix of correct (TP and TN) and incorrect (FP and FN) deletion recommendations (as explained in RQ2), whenever possible. In total, we evaluated 650 UI functionalities, with 325 recommended for deletion by RADIATION and 325 that were not recommended for deletion. Our survey included 42 participants selected via convenience sampling from our social and professional network. For each functionality of the app, three users provided evaluations. Figure 5 displays a sample survey question and the response of one participant specifically for the `org.isoron.uhabits` app.

After familiarizing themselves with their assigned apps for at least 20 minutes, we presented a specific feature of the app they had studied and requested that they rate their liking of the feature on a five-point Likert scale. Furthermore, we also asked the participants to express their emotions if the feature were to be removed. We used conventional sentiment scores [13] for evaluation, with −2 indicating strong dislike, 0 indicating neutrality, and +2 indicating strong liking.

## VI. CASE STUDY RESULTS

Table III presents the results of **RQ1** and **RQ2** for 25 apps that were cross-validated and evaluated by developers. Figure 6 demonstrates the goodness of the topic modeling of app reviews (Step 5) as part of **RQ2**).

### A. Internal Validation of RADIATION (RQ1)

We conducted cross-validation on 115 apps, 3,364 releases, and a total of 190,062 reviews. The results indicate high precision (*0.83*) and recall of *0.48* using 10-fold cross-validation. The precision is considerably higher than recall because in RADIATION the number of false positives (FP) is much lower than false negatives (FN). In other words, in mobile apps, there have been features that were deleted, but RADIATION is unable to recommend them for deletion (FN) This results in a low recall. RADIATION cannot (and is not designed to) capture all deletions that happen within a mobile app. However, as the first study looking into functionality deletion, we could predict with 83% precision. For several of these "false negatives", we did not find reviews related to an element that has been deleted. Hence, we concluded that the feature would not be deleted, and there were other reasons than user reviews for deleting the UI element. Table III details the confusion matrix for the 25 apps that were also externally evaluated in **RQ2**.

> RADIATION *demonstrates 83% precision in recommending deletions based on user reviews. The low recall indicates that not all deletions in a mobile app are motivated by user reviews, which* RADIATION *is not designed to capture.*

### B. External Evaluation of RADIATION with Developers (RQ2)

The 37 developers evaluated RADIATION in two steps.

TABLE III
EVALUATING RESULTS BY COMPARING RADIATION RECOMMENDATIONS WITH (I) RETROSPECTIVE ANALYSIS OF ACTUAL DELETIONS AND (II) DEVELOPERS' PERCEPTION. ONE USER REVIEW MIGHT BE RELEVANT TO MULTIPLE ELEMENTS.

| App's package name | # of UI element across releases | # of reviews | Actual deletions (RQ1) | | | | | Developers' perception (RQ2) | | | | |
|---|---|---|---|---|---|---|---|---|---|---|---|---|
| | | | # of FP | # of FN | # of TP | # of TN | F1 score | # of FP | # of FN | # of TP | # of TN | F1 score |
| (A1) app.openconnect | 235 | 232 | 0 | 2 | 1 | 232 | 0.5 | 0 | 0 | 1 | 234 | 1 |
| (A2) com.google.android.stardroid | 1603 | 4480 | 0 | 2 | 18 | 1583 | 0.95 | 1 | 2 | 18 | 1582 | 0.92 |
| (A3) com.moez.QKSMS | 3009 | 2751 | 0 | 11 | 5 | 2993 | 0.48 | 2 | 4 | 5 | 2998 | 0.62 |
| (A4) com.vuze.android.remote | 774 | 494 | 0 | 2 | 8 | 764 | 0.89 | 1 | 0 | 7 | 766 | 0.93 |
| (A5) net.nurik.roman.muzei | 1088 | 4481 | 0 | 15 | 36 | 1037 | 0.83 | 1 | 0 | 35 | 1052 | 0.99 |
| (A6) org.androisoft.app.permision | 189 | 397 | 0 | 1 | 2 | 186 | 0.8 | 0 | 1 | 2 | 186 | 0.8 |
| (A7) org.connectbot | 471 | 4493 | 0 | 6 | 8 | 457 | 0.73 | 0 | 0 | 8 | 463 | 1 |
| (A8) org.dmfs.tasks | 862 | 207 | 0 | 7 | 7 | 848 | 0.67 | 0 | 4 | 7 | 851 | 0.78 |
| (A9) org.evilsoft.pathfnder.rference | 652 | 1520 | 0 | 0 | 2 | 650 | 1 | 1 | 0 | 1 | 650 | 0.67 |
| (A10) org.isoron.uhabits | 895 | 1976 | 3 | 31 | 101 | 760 | 0.86 | 4 | 13 | 100 | 778 | 0.92 |
| (A11) com.spazedog.mounts2sd | 394 | 497 | 3 | 7 | 60 | 324 | 0.92 | 2 | 0 | 61 | 331 | 0.98 |
| (A12) org.telegram.messenger | 840 | 73682 | 2 | 30 | 26 | 782 | 0.62 | 3 | 3 | 25 | 809 | 0.89 |
| (A13) in.blogspot.anselbros.torchie | 134 | 473 | 8 | 12 | 72 | 42 | 0.88 | 5 | 1 | 75 | 53 | 0.96 |
| (A14) com.emaguy.cleanstatusbar | 86 | 392 | 1 | 7 | 8 | 70 | 0.67 | 0 | 0 | 9 | 77 | 1 |
| (A15) com.boardgamegeek | 1317 | 506 | 33 | 224 | 191 | 6682 | 0.6 | 3 | 25 | 221 | 6881 | 0.94 |
| (A16) com.gelakinetic.mtgfam | 4510 | 2366 | 1 | 3 | 4 | 4502 | 0.67 | 0 | 1 | 5 | 4504 | 0.91 |
| (A17) org.addhen.smssync | 235 | 41 | 6 | 0 | 22 | 207 | 0.88 | 7 | 2 | 21 | 205 | 0.82 |
| (A18) com.amaze.filemanager | 620 | 1241 | 7 | 12 | 25 | 576 | 0.72 | 0 | 1 | 33 | 586 | 0.98 |
| (A19) com.gh4a | 344 | 301 | 4 | 8 | 14 | 318 | 0.7 | 1 | 1 | 17 | 325 | 0.94 |
| (A20) org.kontalk | 54 | 39 | 2 | 2 | 7 | 43 | 0.78 | 2 | 1 | 7 | 44 | 0.82 |
| (A21) org.transdroid.lite | 942 | 538 | 2 | 3 | 7 | 930 | 0.74 | 0 | 0 | 9 | 933 | 1 |
| (A22) de.qspool.clementineremote | 444 | 355 | 4 | 9 | 13 | 418 | 0.67 | 2 | 3 | 15 | 424 | 0.86 |
| (A23) com.daiancorp.mh4udtabase | 3101 | 979 | 29 | 51 | 73 | 2948 | 0.65 | 12 | 5 | 90 | 2994 | 0.91 |
| (A24) org.servalproject | 547 | 252 | 4 | 14 | 10 | 519 | 0.53 | 2 | 3 | 15 | 527 | 0.85 |
| (A25) org.wikipedia | 17830 | 15531 | 23 | 0 | 94 | 17713 | 0.89 | 1 | 0 | 116 | 17713 | 0.99 |
| **Average** | 1647.04 | 4728.96 | 5.28 | 18.36 | 32.56 | 1823.36 | 0.74 | 2 | 2.8 | 36 | 1838.84 | 0.9 |

**FP** (False-Positive): Recommended as deletion but was not, **FN** (False-Negative): Recommended not a deletion but it is, **TP** (True-Positive): Recommended as deletion and it is, **TN (True-Negative)**: Recommended as not a deletion and is not.

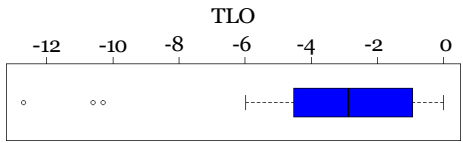

Fig. 6. Topic Log Odds (TLO) shows the performance of RADIATION's clustering against developers' perception.

*1) Evaluation of cluster topics about each UI element:* We followed the approach of Palomba et al. [31] to cluster user reviews by their connection to UI elements. Hence, in RADIATION we first connected reviews to the UI elements (Step 4 ) and then clustered the reviews around each UI element using HDP topic modeling (Step 5 ) [31]. We presented the number of UI elements along with the number of clusters and number of user reviews in Table III. To evaluate the usefulness of our topic model, we relied on the judgment of app developers. After asking them to evaluate the topics using topic intrusion, we calculated TLO as suggested by Chang et al. [4]. We present the distribution of TLO in the boxplot chart of Figure 6. $TLO = 0$ shows the highest conformance between developers and the topic modeling technique. Comparison of the distribution of our HDP clustering showed a slight disagreement between developers and machine learning results as the median is around $-3$. However, this is still considered as a relatively low disagreement compared to former benchmarks [3], [4].

*2) Evaluating RADIATION recommendations:* We asked developers to evaluate whether a cluster of reviews for a UI element were "motivating a functionality deletion" or "not motivating a functionality deletion" (e.g., implying a bug fix). We compared RADIATION results to developer perceptions for 25 randomly selected apps, resulting in an average F-Score of 90% for RADIATION. See Table III for the number of true and false recommendations for these apps.

Upon examining the results presented in Table III, it is apparent that there are fewer false positives (FP) and false negatives (FN) when comparing our recommendations with developers' perceptions as opposed to retrospective evaluation. This difference can be attributed to the fact that recommending deletions involves multiple factors beyond user reviews, which RADIATION does not take into account. Therefore, when asking developers to make a decision based on user reviews, RADIATION demonstrates better performance.

> RADIATION *achieves an average F-score of 0.9 when its recommendations are compared with the developers' decisions based on the respective clustered reviews.*

### C. External evaluation of RADIATION with Users (RQ3)

Our objective was to evaluate user sentiment towards the functionalities recommended for deletion by RADIATION. To achieve this, we conducted a survey of 42 users to evaluate their perception of specific mobile app functionalities and to understand their sentiments if those functionalities were to be removed (refer to Figure 5). We asked each participant two questions regarding the features they were evaluating. Figure

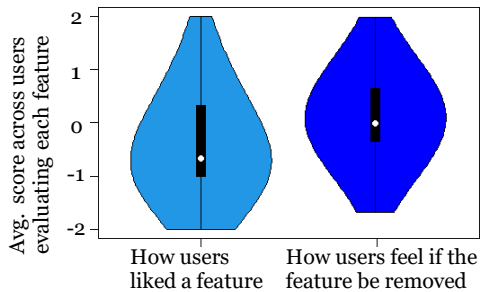

Fig. 7. Evaluation results of 650 features with users through survey

2 displays a violin plot of the results. Table IV presents an overview of our results for the first survey question we asked in **RQ3** and for each of the 25 apps we evaluated. Each column in this table displays the average of responses provided by three users who participated in our survey. Note that the number of samples was not evenly distributed across TP, TN, and other categories. For example, (A1) app.openconnect had only one UI functionality that was correctly recommended for deletion (TP) in **RQ2** (see Table III). We also asked users how they would feel if the functionality were to be removed (Q2). We observed a high correlation of -0.86 between the responses to Q1 and Q2 in our survey. That being said, we found that the more negative the feelings users had towards the feature, the more positive they were about its removal.

When we surveyed users about the functionalities, we observed that the average sentiment of the participants towards the features that were correctly recommended for deletion by RADIATION (TP recommendations) was consistently negative. In other words, the negative experiences of the users were aligned with the recommendations. However, for deletions that were not actually performed (FP), we observed mixed sentiments. Nevertheless, the majority of the apps (13 out of 16) received an overall average of negative sentiments for wrong predictions as well. Thus, it is essential to note that a negative experience might not necessarily imply feature deletion but could call for a bug fix or a change in the software. This finding aligns with our analysis of **RQ2**, where external developers favored RADIATION recommendations, while historical data showed that the decisions of the actual app developers (**RQ1**) were different. This difference could be due to the exclusion of particular ecosystem or business factors in RADIATION modeling.

> *The users consistently disliked the functionalities that* RADIATION *correctly recommended for deletion and in general are not against removing them.*

## VII. DISCUSSION

In this section, we briefly discuss the further interpretation of the achieved results and some design decisions.

### A. Scope of RADIATION

Motivated by the number of studies on release planning of mobile applications and in consideration of the limited resources for mobile devices [26] we studied the possibility of predicting feature deletions for mobile applications. RADIATION uses user reviews to recommend UI functionality deletions based on various factors. We analyzed user reviews and clustered them according to relevant UI elements, which enables RADIATION to focus solely on user feedback and visible app functionality. Upon retrospective analysis, we found that RADIATION has a low recall due to a considerable proportion of false negatives. These false negatives indicate deletions that were not motivated by user reviews and therefore fell outside the scope of RADIATION recommendations. To further evaluate the effectiveness of our approach, we provided software developers with reviews for each UI element and asked them to decide whether they motivated functionality deletion or not. This resulted in better recall compared to our previous cross-validation results. We also evaluated user sentiment toward these functionalities and found that they consistently experienced negative emotions when using the RADIATION recommended for deletion. We further discovered that the more negative the user's experience, the more likely they were to be neutral or positive about removing that feature from the app.

### B. Benchmarking and performance of RADIATION

We relied on the highly performed methods discussed in the literature and did not re-evaluate the performance of the learners. We do not argue these techniques are the most optimal and highest-performing methods possible. Rather, as the first study on recommending feature deletion in app releases, we focused on exploring the possibility of deletion recommendations, their usefulness, and the ease of explanation to the users and the developers. As the first study on predicting deletions based on user reviews, our target was to examine if the deletion prediction is possible rather than to highly optimize the performance of the approach. This is essential step before taking further steps for planning these deletion. Based on the current state-of-the-art results, we do not expect that a benchmark of different classifiers would significantly improve the performance of our approach.

One key motivation for the paper comes from the observation that current release planning in general [36] and in particular for mobile apps [20], [44] is exclusively focused on feature addition. Planning in consideration of both addition and deletion of functionality requires revisiting the planning objective(s). Clearly, deletion consumes development effort as well. While we took the first step toward understanding functionality deletion, future work involves contextualizing the results for specific projects and development teams. Besides a more comprehensive empirical evaluation in general, we also target trade-off analysis between measuring the evolving maintenance effort and functionality deletions. Overall, the main goal of future research will be to better understand the deletion of functionality as part of software evolution, also beyond mobile apps. In addition, we will work on improving the performance of our recommendations by updating the machine learning techniques and features and tuning the model (for instance, by more in-depth analysis of similarity).

## VIII. THREATS TO VALIDITY

Throughout the different steps of the process, there are various threats to the validity of our achieved results.

**Are we measuring the right things?** We pre-processed all review texts and used machine learning classification to ensure that the analysis is only considering informative user reviews. The Naive Bayes classification resulted in an F1 score of 0.82. While this is a very good result, there is still a possibility that a review has been classified incorrectly. There is a risk related to linking reviews to the proper UI elements. Two of the authors looked into the results of this linking (Step 4 of RADIATION) for 600 reviews across six apps and found 71 mismatched or unrelated reviews.

**Are we drawing the right conclusions about treatment and outcome relation?** In comparison to studies in the context of mobile apps (Table V), our surveys can be considered high participated. However we used convenience sampling to attract participant which might bias the conclusions that are drawn [16]. It is essential to note this type of evaluation is subjective. However, the results of **RQ1** based on the retrospective analysis of the data are aligned with our survey results presented in **RQ2** and **RQ3**. In total, we think that the evaluation gained with 37 developers and 42 users is sufficient to confirm our findings.

When connecting a review to a UI element in RADIATION, there is a chance that we relate a review to an element incorrectly (false positives). This may happen because
- We may miss some UI elements, as they can be instantiated in the program code or hard coded,
- Some UI elements are not visible to the end user, or
- Text of some UI elements are common English words or can have similar labels in different app views.

TABLE IV
EVALUATING USER SENTIMENTS TOWARD THE FEATURES RADIATION RECOMMENDS FOR DELETION THROUGH A SURVEY (**RQ3**)

| App ID | Q1: Average Sentiment toward functionalities that are | | |
|---|---|---|---|
| | Incorrect deletion recom. (FP) | Correct deletion recom. (TP) | other (FN or TN) |
| (A1) | N/A | -1.3 | 2.0 |
| (A2) | -0.13 | -1.07 | 0.86 |
| (A3) | -0.7 | -1.16 | 1.0 |
| (A4) | -0.55 | -0.66 | -0.66 |
| (A5) | 1.07 | -1.0 | 0.08 |
| (A6) | N/A | -0.86 | -1.13 |
| (A7) | N/A | -0.93 | 0.0 |
| (A8) | N/A | -0.06 | 0.13 |
| (A9) | -0.66 | -1.2 | 0.91 |
| (A10) | 1.27 | -0.91 | 1.13 |
| (A11) | -0.55 | -1.0 | 1.05 |
| (A12) | -0.45 | -1.8 | -0.79 |
| (A13) | -0.56 | -1.0 | 0.51 |
| (A14) | N/A | -2.0 | 0.06 |
| (A15) | -0.88 | -1.4 | 0.79 |
| (A16) | N/A | -1.66 | 0.81 |
| (A17) | -0.77 | -1.0 | 0.73 |
| (A18) | N/A | -1.66 | 0.21 |
| (A19) | 0.97 | -1.08 | 0.91 |
| (A20) | 0 | -1.16 | 1.21 |
| (A21) | N/A | -1.13 | 0.05 |
| (A22) | -1.03 | -1.5 | -0.31 |
| (A23) | -0.89 | -1.33 | -1.09 |
| (A24) | -0.09 | -1.55 | -0.45 |
| (A25) | -1.02 | -1.02 | 0.18 |

TABLE V
CONTEXT AND EVALUATION OF RELATED STUDIES.

| Method | Context | Evaluation |
|---|---|---|
| ARdoc [33] | Information giving/seeking, feature request, problem discovery, others | Evaluating three apps by two developers |
| AR-Miner [5] | Informative or non-informative reviews | Manual inspection by authors, comparison between techniques |
| CHANGEADVISOR [31] | Localizing change request by linking reviews and source code | Evaluated results with 12 developers |
| CLAP [44] | New feature request, bug report | Retrospective analysis of 463 reviews and interview with three developers |
| CRISTAL [29] | Tracing user reviews to the developers changes | Manual evaluation by authors |
| MARA [14] | Feature request | Comparing different techniques |
| PAID [9] | Issues (bugs) | Retrospective analysis of 18 apps |
| Panichella et al. [32] | Information giving/seeking, feature request, problem discovery, others | Comparison between different methods |
| SURF [7], [8] | Information giving/seeking, feature request, problem discovery, others | 23 developers analyzed SURF output for 2622 reviews. |
| SUR-Miner [12] | Aspect evaluation, praise, function request, bug report, others | Comparing techniques, evaluation with 32 developers |
| URR [6] | Compatibility, usage, resources, pricing, protection, complaint | Qualitative evaluation by a student and a developer |

To address the first two items above, we used BACKSTAGE [1] on a few of the apps and we found that while the risk exists, it is relatively small. Since BACKSTAGE works on compiled application binaries we were limited to using it in RADIATION. For the third item above, we applied preprocessing as suggested in CRISTAL [29] and adopted their list of stop words. Further, RADIATION is not intended to exhaustively find all the deleted feature (recall). The impact of potentially missed elements is insignificant.

**Can we be sure that the treatment indeed caused the outcome?** The selection of attributes used in RADIATION to decide *if a UI functionality should be deleted* is another threat to validity. Our survey with users was aligned with the findings in the literature [26] and showed that users and their feedback is important information in the deletion process. However, it is not the only decisive factor for excluding a functionality from apps. We selected attributes based on related studies (Table II). There are other attributes related to competitors, performance, or maintenance considerations that are relevant for the decision-making but could not be taken into account for our study. Following the results of former studies on mobile apps [29], we assumed that users are reviewing just the functionality that is visible to them (and not the background code). This might not be true for all the users, reviews, and sentiments. However, we expect a low number of such cases.

**Can the results be generalized beyond the scope of this study?** Our retrospective analysis was performed on open-source mobile apps. The number of apps, reviews, and commits analyzed is considered high, indicating that results are significant at least for open-source mobile apps. While selecting the apps for this study, we did not consider their status (for example, the number of downloads) which may pose a risk of bias in the findings. The results may vary between apps with regards to their status on the app store[3].

## IX. RELATED WORK

In this study, we challenged Lehman's law of growth by investigating functionality deletion as a specific activity in the development process. We focused on the mobile apps because the device resources are limited and the size of the release has been introduced as a decisive factor for release decisions [22], [27], [28]. Feature and functionality deletion for software products in general have been discussed mostly on the model level which triggered us to widely investigate on the nature and reasons of functionality deletion in **RQ1** and **RQ2**.

[3]Authors will provide data and scripts in case of acceptance.

Analyzing user reviews to support app evolution and maintenance was studied by several researchers [20]. These studies are mainly focused on different user needs to be articulated at the level of being a "feature request" or "bug report" [18]. The study by Palomba et al. [29] found that 49% of informative reviews were considered for app evolution. In this direction, current studies take user reviews as the source of change requests, apply a variety of NLP techniques, and provide a prioritization or classification scheme. The objective is to help developers decide on the next best changes either by adding new functionality or fixing a bug. We provided an overview of the most related methods in Table V.

> *Current literature discuss different types of user requests on app evolution. We focused on a functionality deletions which was not studied.*

CLAP [44] used a mixed method by combining the retrospective analysis of changes for 463 reviews in conjunction with interviewing three app developers. PAID [9] had the most comprehensive retrospective evaluation of data by investigating 18 apps for issue (bug) prioritization. Compared to the former studies in analyzing app reviews, we have a more rigorous evaluation by asking 37 developers to evaluate 36,039 reviews for a total of 25 apps. We compared these evaluations with the results gained from RADIATION. While some studies compared different methods for evaluating their results, this was not possible for RADIATION in general as none of the existing techniques is focused on functionality deletion. However, to select classifier and topic modeling techniques, we made the comparisons as discussed in Section IV.

## X. CONCLUSIONS

*Lehman's law on continuous growth of functionality does not universally apply.* In the domain of mobile apps, developers frequently delete functionality—be it to fix bugs, maintain compatibility, or improve the user experience. We performed a study with *app users* to confirm the potential value of deletions also from their perspective. We suggested that the process of selecting the functionality to be deleted can be automated, as demonstrated by our RADIATION recommendation system. RADIATION analyses the UI elements of the app and the reviews and recommends if the UI element and its functionality shall be deleted or not. This is the first study to investigate the prediction of functionality deletion in software evolution. It opens the door towards a better understanding of software evolution, in particular in an important domain such as mobile app development. In the days of Lehman's studies, features such as user experience, screen space, or energy consumption were not as crucial as they are today; it may be time to revisit and refine Lehman's findings.